# Understanding the negative temperature coefficient phenomenon in methane-air mixtures at high pressures

Anand Parejiya[a], Manjeet Chaudhary[a], Sai Mani Prudhvi Valleti[a], Marm Dixit[a], Atul Bhargav[a,1], Suman Roy Choudhury[b]

[a]Energy Systems Research Laboratory, IIT Gandhinagar, Ahmedabad, Gujarat, INDIA

[b]Fuel Cell Research Center, Naval Materials Research Laboratory, Ambarnath, Maharashtra, INDIA

## Abstract

Design and operation of advanced reactors such as fuel reformers require reliable micro-kinetic models that capture the dynamics of the reaction. The negative temperature coefficient phenomenon causes a reduction in mixture temperature for increasing inlet temperatures. However, micro-kinetic models available in the literature have not been critically evaluated for their ability to capture this phenomenon. Consequently, the ability to predict system behavior for particular application situations, such as in the presence of certain diluents or at high pressures, is largely missing. In this work, we adapt multiple reaction mechanisms from literature and compare them for methane oxidation over a wide range of pressures and temperatures. Using reaction path analysis and sensitivity analysis, we find that the C2 formation through the recombination pathway is chiefly responsible for negative temperature coefficient behaviour. With this insight, the dependence of steam addition and pressure on is also discussed.

[1] Corresponding author: Dr. Atul Bhargav, Assistant Professor (Mechanical Engineering)
A: IIT Gandhinagar (VGEC Campus), Visat-Gandhinagar Highway, Chandkheda, Ahmedabad, GJ 382424 INDIA
E: atul.bhargav@iitgn.ac.in | P: +91 814 030 7813

| Symbol | Description | Units |
|---|---|---|
| $\beta$ | temperature dependence factor | - |
| $\lambda$ | excess air ratio | - |
| $\psi$ | dilution factor | - |
| $\tau_{ign}$ | auto ignition delay | s |
| $\rho$ | density | kg m$^{-3}$ |
| $\dot{\omega}_k$ | Net volumetric production rate of species '$k$' | kmol m$^{-3}$ s$^{-1}$ |
| $A$ | pre-exponential factor | s$^{-1}$ |
| $c_p$ | specific heat | kJ kg$^{-1}$ K$^{-1}$ |
| $E$ | activation energy | kJ mol$^{-1}$ |
| $h_k$ | enthalpy of species '$k$' | kJ/mol |
| $K$ | total number of species | - |
| $k_f$ | rate constant | reaction specific |
| $N$ | total number of reactions | - |
| $\dot{r}_j$ | volumetric reaction rate of reaction number '$j$' | mol cm$^{-3}$ s$^{-1}$ |
| R | universal gas constant | kJ mol$^{-1}$ K$^{-1}$ |
| $T$ | temperature | K |
| $v_{j,k}$ | stoichiometric co-efficient of species '$k$' for reaction '$j$' | - |
| $W_k$ | molecular weight of species '$k$' | kg kmol$^{-1}$ |
| $Y_k$ | mole fraction of species '$k$' | - |

Acronyms/ Abbreviations

| | | |
|---|---|---|
| NTC | negative temperature coefficient | |

# 1    Introduction

Development of alternatives to conventional practices of utilization of fossil fuel sources is a major research focus around the world to curb the increasing pollution and global warming effects as well as increasing energy efficieny (Seidel et al. 2015; Ameen et al. 2015; Cheng et al. 2015; Prince et al. 2015; Battin-Leclerc 2008; Zádor et al. 2011).. In particular, owing to the large reserves of natural gas(U.S. Energy Information Administration 2014) , there have been substantial experimental and numerical studies involving methane, the primary component of natural gas (Diamantis et al. 2015; Hoang & Chan 2004; Wehinger et al. 2015; Huang et al. 2004; Dixit et al. 2015; Fukada 2004; Ranzi et al. 2007; Deshmukh & Vlachos 2007; Wei 2004). Some recent efforts have been focused on experimental and numerical studies of methane-oxygen mixtures, with a specific focus on MILD combustion regimes (Sabia et al. 2013). Further, negative temperature co-efficient (NTC) regime has been experimentally verified for methane recently (Picarelli et al. 2010; Sabia et al. 2013). Typically, this regime is characterized by moderate to intensely low-oxygen dilution and is finding application in numerous real-life applications (Sabia et al. 2015). The study of methane oxidation is expected to find application in industrial furnaces, reformers, automotive engines as well as solid-oxide fuel cell systems (Vourliotakis et al. 2009; Zhou et al. 2010; Belmont & Ellzey 2014; Xu et al. 2014; Gallagher et al. 2008; Seidel et al. 2015; Vourliotakis et al. 2012). In addition, from a safety perspective, there is a need to study the NTC region of oxy-fuel mixtures (Pekalski et al. 2002).

This behavior was previously known for higher hydrocarbons (Naidja 2003; Hartmann et al. 2003), and limted literature exists for methane oxidation. However, even this available literature limited to experiments carried out at relatively low temperatures and pressures (Healy et al. 2008; Sabia et al. 2013). Picarelli et al studied diluted methane air mixtures numerically and experimentally (Picarelli et al. 2010). Sabia et al experimentally studied the NTC regimes for diluted methane-oxygen mixtures at intermediate temperatures and compared multiple reaction mechanisms for their ability to predict the same (Sabia et al. 2012; Sabia et al. 2013). However these studies do not quantify the effects of additional co-reactants like steam, as well as the effect of pressure on the system response. Steam is an important reactant in reforming and pre-reforming processes, where the NTC region is observed (Hartmann et al. 2003). We have employed multiple mechanisms available in open literature and compared them for methane oxidation at higher pressures. Sensitivity analysis and reaction path analysis have been performed to understand the important reactions and pathways.

The analyses shown in this paper has significance both in terms of the process as well as the numerical tools available to simulate these systems. Because the methane-oxidation system is a multi equilibrium system, various contending pathways are possible that affect the kinetics of the reaction. Therefore, the interdependence of working temperature, reactant concentration of and the various branches of the active oxidation pathway were studied. Section two details the various mechanisms adapted from literature and details the numerical model investigated. Section three describes the results for simulations and effect of reactant addition. Section four details the reaction path analysis and sensitivity analysis performed and elicits the key reactions and pathways.

## 2    Reaction Mechanisms, Numerical Model and Validation

Numerical simulations for studying the evolution of methane ignition process were carried out using the thermo-kinetic calculator toolbox "Cantera" (Goodwin et al. 2015). Several oxidation kinetic mechanisms available in the literature were considered. The names used to identify them in this paper, along with details of number of involved species and reactions are summarized in Table 1. Kinetic mechanisms employed here are extensively validated for several operating conditions and reactor configurations in literature (Ranzi et al. 2007).

**Table 1**: Details of Reaction Mechanism Studied

| Model | Reference | # of species | # of reactions |
|-------|-----------|--------------|----------------|
| GRI-Mech 3.0 | (Smith et al. n.d.) | 53 | 325 |
| Ranzi | (Ranzi et al. 2010) | 107 | 2642 |
| Huang | (Huang et al. 2004) | 40 | 194 |

In the present work, auto-ignition delay times $\tau_{ign}$ were computed from the axial temperature profiles in an adiabatic isobaric reactor for a wide range of parameters. The energy conservation equation is shown in (1).

$$c_p \cdot \frac{dT}{dt} + \frac{1}{\rho} \cdot \sum_{k=1}^{K} h_k \cdot \dot{\omega}_k \cdot W_k = 0 \qquad \text{(a)}$$

Where the mean specific heat of mixture is given by $c_p = \sum_{k=1}^{K} Y_k . c_{p,k}$ . The net chemical production rate

$\dot{\omega}_k = \sum_{j=1}^{N} \nu_{j,k} \cdot \dot{r}_j$ .

$$k_f = A T^{\beta} \exp\left(\frac{-E}{RT}\right), \qquad \text{(b)}$$

In particular, the investigated temperature range is 900–1200 K for an air ratio (λ) of 1.25 and inert gas dilution ($\psi$) of 85% at atmospheric pressure, 5 bar and 10 bar. Following a criterion described in details elsewhere (Joannon et al. 2002), auto ignition delay $\tau_{ign}$ is defined as the time corresponding to a temperature increase of 10 K with respect to reactor inlet temperature ($T_{in}$). In addition to auto-ignition delay computation, CANTERA application has also been used for reaction sensitivity and rate of production analyses and automatically drawing net reaction path diagrams.

# 3   Results and Discussions

The $\tau_{ign}$ computed for an 85% diluted mixture with an excess air ratio of 1.25 obtained by the various mechanisms is shown in . As can be seen, auto-ignition delay decreases monotonically with increase in inlet temperature. However, a large variation is seen in the predicted delays. For instance, at 900°C, the ignition delay predicted by the Huang mechanism is an order of magnitude less than that predicted by GRI 30. Also, while the predicted delay by GRI 30 is linear with respect to inlet temperature, the other two mechanisms show a slight deviation from this behavior.

Between 950 K and 1100 K the Huang and Ranzi mechanisms predict- a change of slope. This behavior can be explained by a change in the system reactivity. In particular, the Ranzi mechanism has $\tau_{ign}$= 0.851 s at 950 °C, and it drops only to 0.546 s at 1000°C; this amounts to at least two orders of magnitude larger time delay change than that expected of a system with linear response. The Huang mechanism shows a much smoother transition between the two linear regimes as compared to Ranzi mechanism.  This behavior has not been characterized extensively experimentally, and hence the exact reason for this behavior is not known. However, it is clear that certain kinetic pathways in the intermediate temperature regime are responsible for this behavior.

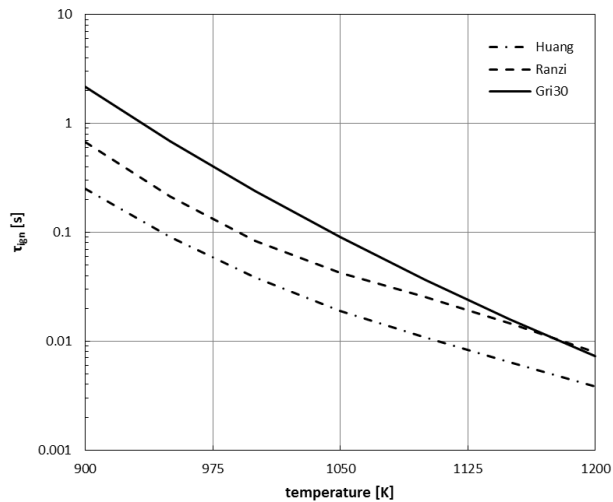

**Figure 1:** $\tau_{ign}$ for various reaction mechanisms and inlet temperatures at $\lambda$ = 1.25 and $\psi$ = 85%

Effect of pressure on the ignition delay has also been studied Figure 2. The increase in pressure yields a corresponding decrease in the auto-ignition delay. This behaviour is expected as an increase in pressure increases the reactivity of the system by allowing increased collisions between the reactant molecules. We also observe that the NTC region shifts to higher inlet temperatures with increase in pressure. With competing pathways and increased reactivity, this shift has also been seen experimentally (Healy et al. 2010). NTC behavior disappears at 10 bar for the temperature range examined for Huang mechanism, while at higher pressures the Ranzi mechanism shows a continuous non-linear behavior.

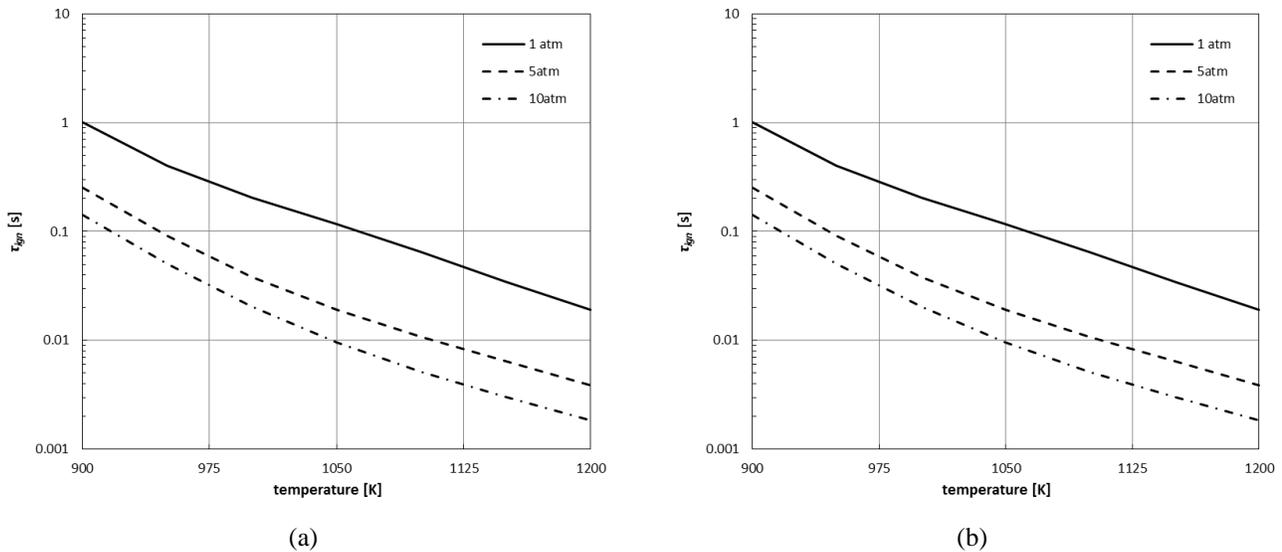

(a)                                        (b)

**Figure 2:** Effect of pressure on $\tau_{ign}$ at $\lambda$ = 1.25 and $\psi$ = 85% for: (a) Huang mechanism (b) Ranzi mechanism

Effect of steam addition on the ignition delay has also been studied. As can be seen from , steam addition increases the auto ignition delay across all the temperatures and pressures studied, which is to be expected.

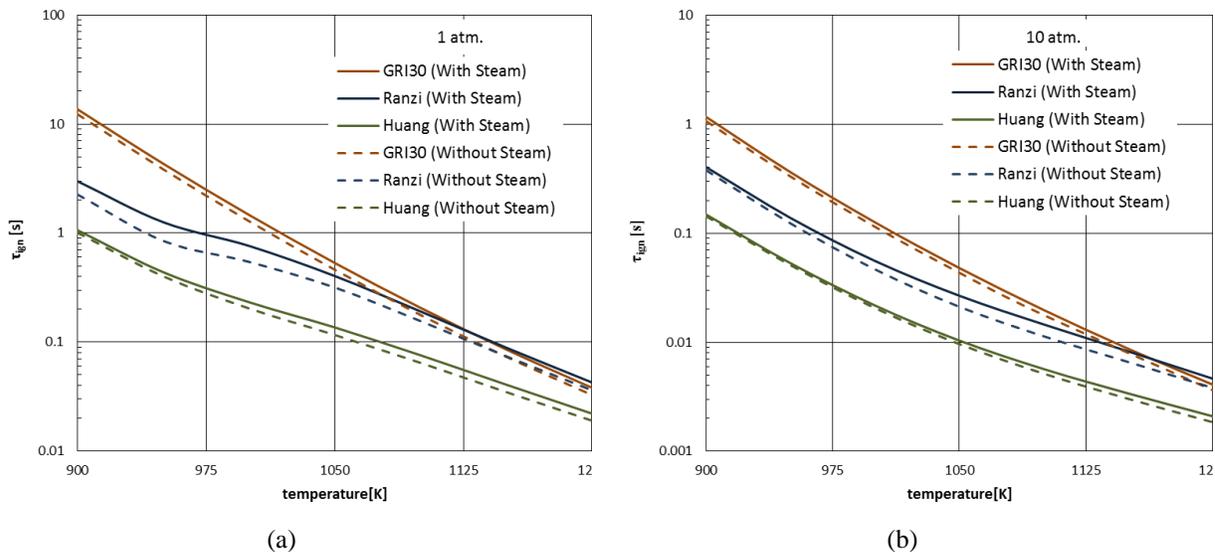

(a)          (b)

Figure 3 Effect of steam addition on $\tau_{ign}$ at $\lambda$ = 1.25 and $\psi$ = 85% for: (a) P = 1 atm. (b) P = 10 atm.

## 4    Reaction Path Diagram and Sensitivity Analysis

Reaction path analysis (RPA) helps identifying the main reaction paths from reactants to products, and is based on the net production rates of surface species involved in the microkinetic mechanism (Mhadeshwar & Vlachos 2005). Since the Ranzi mechanism shows the best behavior, RPA and sensitivity analysis have been carried out for this Ranzi mechanism using the "object oriented programming tool" within Cantera. A reaction path diagram for reactor data described in previous section is shown in Figure 4. At 900 K, three competing pathways exist lead to CO and $CO_2$: the major pathway is $CH_3OH$ to $CH_2OH$ to folmyl radical. The other chief contending pathway follows a dissociative mechanism with formation of $CH_2S$, $CH_2$ and then formyl radical. Another minor pathway observed is $CH_3$ oxidizing to $CH_3O$, formyl radical and then to CO. However, at higher temperatures, where NTC is observed, a longer pathway is found to be active. Not only does the $CH_3$ radical follow the three pathways discussed previously, it undergoes coupling reaction to form ethane. Ethane then undergoes dissociation to form $CH_2CO$ radical, subsequently yielding $CH_2O$. Then, a similar reaction path as the non-NTC region is observed. A substantial fraction of C follows this latter pathways leading to a decrease in the reactivity, and conclusively increases the auto-ignition delay. While at low pressures, the dissociative and oxidative pathways with the CH₃O and CHₓ intermediate are favored, at higher pressures the methyl alcohol intermediate is the most abundant. $CH_3OH$ forms from the $CH_3$ radical which then reduces to $CH_2O$. This radical then dissociates to give HCO, the precursor to CO for all contending pathways. This behavior is seen at higher pressures for all temperatures evaluated. Addition of steam reinforces the dissociative pathway for all pressures and temperatures. If we assume the C2 pathway as being responsible for the NTC behavior, the behavior of steam addition on ignition delay can be understood.

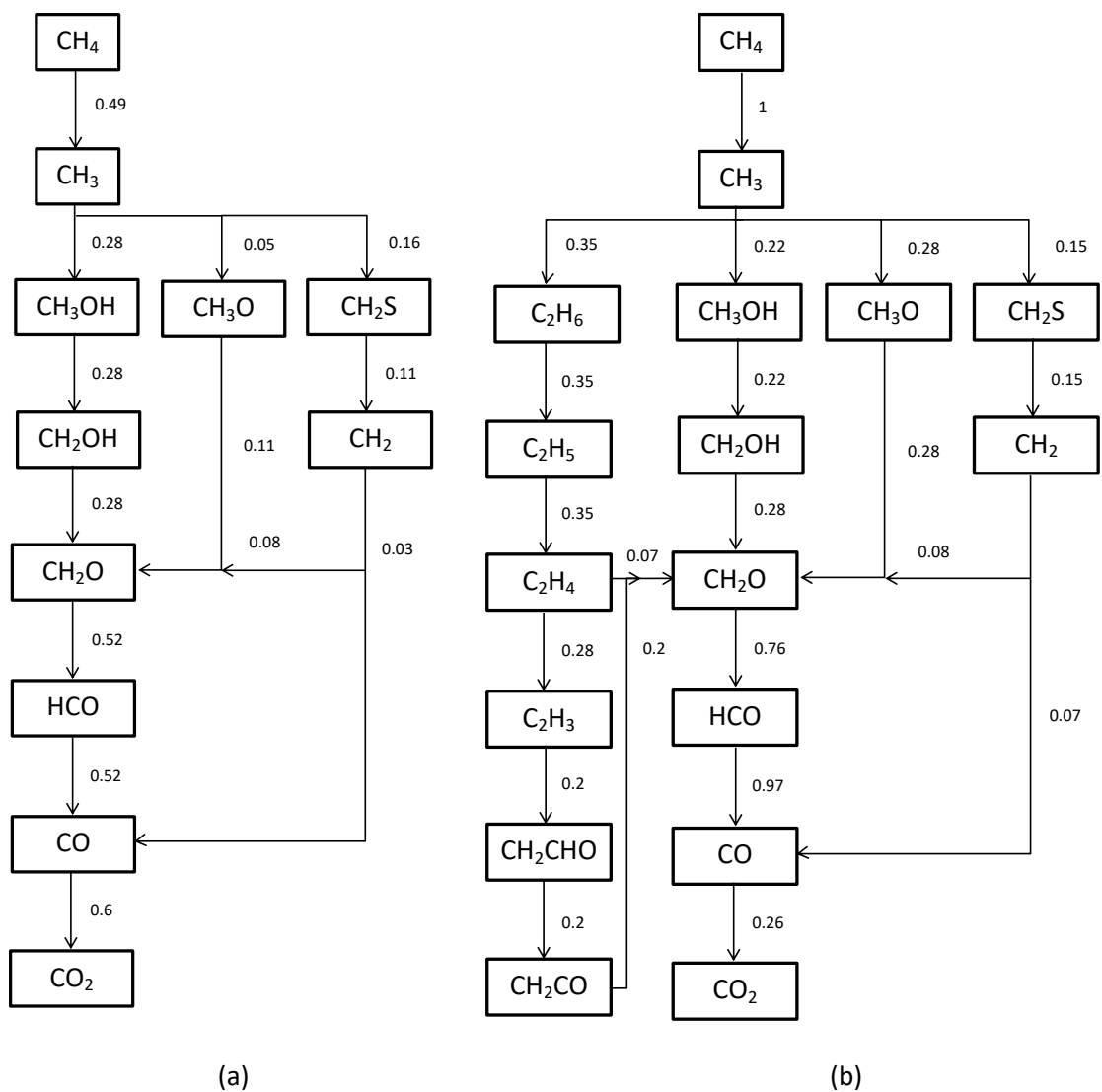

(a)                    (b)

**Figure 4:** Reaction path analysis at the reactor exit (a) Outside NTC region and (b) Inside NTC region. Values are for the fraction of C which proceeds through a given irreversible reaction (may not sum to 100% due to rounding).

Based on the results obtained from RPA, the following key reactions were identified:

| | |
|---|---|
| $H + O_2 + M \rightarrow HO_2 + M$ | 1 |
| $HO_2 + HO_2 \rightarrow H_2O_2 + O_2$ | 2 |
| $H_2O_2 + M \rightarrow OH + OH + M$ | 3 |
| $H + O_2 \rightarrow OH + H$ | 4 |
| $CH_4 + H \rightarrow CH_3 + H_2$ | 5 |
| $CH_4 + HO_2 \rightarrow CH_3 + H_2O_2$ | 6 |
| $CH_4 + O_2 \rightarrow CH_3 + HO_2$ | 7 |
| $CH_3 + HO_2 \rightarrow CH_3O + OH$ | 8 |
| $CH_3 + O_2 \rightarrow CH_3O + O$ | 9 |
| $CH_3 + CH_3O_2 \rightarrow CH_3O + CH_3O$ | 10 |
| $CH_3 + O_2 \rightarrow CH_3O_2$ | 11 |
| $CH_3 + CH_3 + M \rightarrow C_2H_6 + M$ | 12 |
| $CH_3O + M \rightarrow CH_2O + H + M$ | 13 |
| $CH_3O_2 \rightarrow CH_2O + OH$ | 14 |
| $CH_3O_2 + CH_4 \rightarrow CH_3OOH + CH_3$ | 15 |
| $CH_3 + CH_2O => CH_4 + HCO$ | 16 |

Sensitivity analysis was performed on the predicted auto-ignition delay by perturbing the reaction rates of each of the above reaction by 5% at various temperatures and pressures (Figure 5) As can be seen, single carbon pathways are the most sensitive in the temperature ranges outside the NTC regime. The recombination pathways have opposing and large sensitivities in the NTC regime. It is seen that reaction yielding $C_2H_6$ has the highest sensitivity at 1100 K, resulting in reduced $CH_3$ radicals available for the dissociative and oxidative pathways. The increase in auto-ignition delay is due to the decreased system reactivity following the reasons outlined. Further, at higher pressures, this reaction shows reduced sensitivity, which explains the reduced NTC behavior at those pressures. Similar behavior has been observed for the case when steam is introduced. Both the sensitivity analysis and the reaction path diagram corroborate the proposed pathways in the NTC regime. Further, the effects of steam addition and pressure can be clearly distinguished, and the conclusions as described can be drawn.

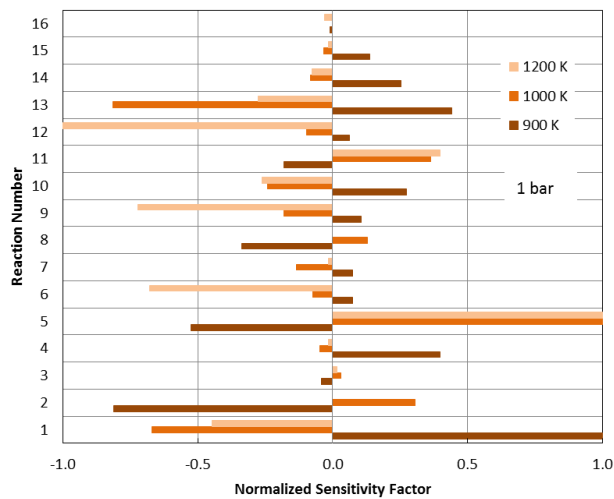

(a)

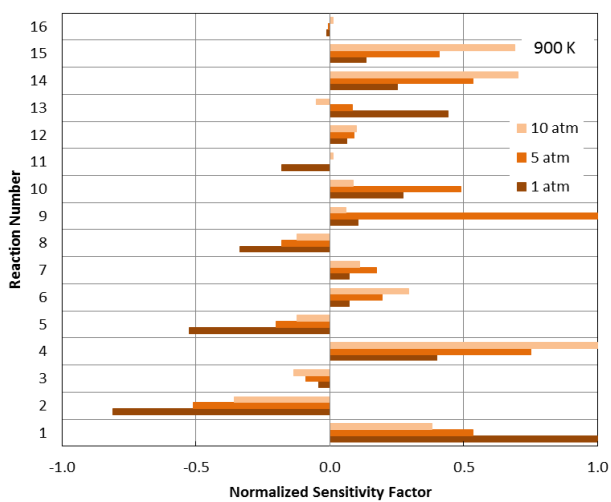

(b)

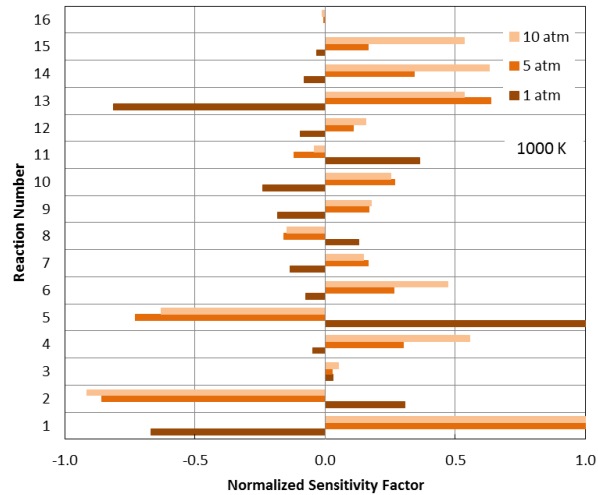

(c)

**Figure 5:** Sensitivity analysis for the Ranzi mechanism: Comparison of sensitivities for the effect of (a) temperature at 900 K, 1000 K and 1200 K (b) pressure at 900 K (c) pressure at 1000 K. Recombination pathway is observed to be activated at interim temperatures; however this effect is subdued at higher pressures.

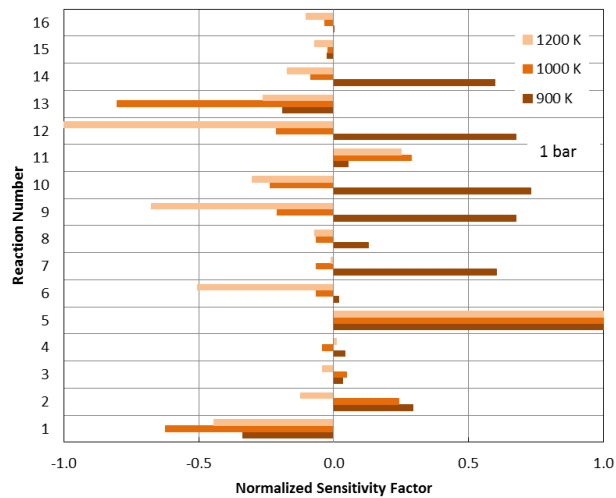

**Figure 6:** Effect of steam addition on the sensitivity analysis for Ranzi mechanism at different temperatures. The system pressure is maintained at 1 bar. The increase in sensitivity of the recombination reaction (reaction number 13) at 1000 K indicate the shift of the reaction pathway.

## 5   Conclusion

The present work attempts to understand the auto-ignition behavior of methane-oxygen mixtures in diluted conditions at high pressures. Several available reaction mechanisms were investigated for their ability to model negative temperature coefficient regime. Specifically, behavior of diluted mixtures was investigated at atmospheric to 10 bar pressures. Effect of steam addition to the reactant mixture was investigated for the entire range of temperatures and pressures. From reaction path analysis and sensitivity analysis it is seen that recombination pathway that yield C2 species slows down the reactive process leading to the NTC behavior. This study also indicates that of the available reaction mechanisms in the literature, only the Ranzi mechanism clearly captures the NTC regime, while there is strong evidence in the experimental literature of NTC's existence. Therefore, it may be necessary to carry out further experimental and modeling studies in the MILD combustion regime.

# 6  Acknowledgement

The authors acknowledge support received from the Ministry of Human Resource Development, Government of India, and the Ministry of Science &Technology, Government of India through grant SR/ S3/CE/078/2012 and Naval Materials Research Laboratory through CARS Project No. NMRL/EST/NMR238/41/2014-15.

# 7  References

Ameen, M.M., Magi, V. & Abraham, J., 2015. An Evaluation of the Assumptions of the Flamelet Model for Diesel Combustion Modeling. *Chemical Engineering Science*, 138, pp.403–413. Available at: http://linkinghub.elsevier.com/retrieve/pii/S0009250915005898 [Accessed August 31, 2015].

Battin-Leclerc, F., 2008. Detailed chemical kinetic models for the low-temperature combustion of hydrocarbons with application to gasoline and diesel fuel surrogates. *Progress in Energy and Combustion Science*, 34(4), pp.440–498. Available at: http://linkinghub.elsevier.com/retrieve/pii/S0360128507000627 [Accessed September 18, 2015].

Belmont, E.L. & Ellzey, J.L., 2014. Lean heptane and propane combustion in a non-catalytic parallel-plate counter-flow reactor. *Combustion and Flame*, 161(4), pp.1055–1062. Available at: http://www.sciencedirect.com/science/article/pii/S0010218013004008 [Accessed January 1, 2015].

Cheng, X. et al., 2015. Development and validation of a generic reduced chemical kinetic mechanism for CFD spray combustion modelling of biodiesel fuels. *Combustion and Flame*, 162(6), pp.2354–2370.

Available at: http://linkinghub.elsevier.com/retrieve/pii/S0010218015000449 [Accessed July 27, 2015].

Deshmukh, S. & Vlachos, D., 2007. A reduced mechanism for methane and one-step rate expressions for fuel-lean catalytic combustion of small alkanes on noble metals. *Combustion and Flame*, 149(4), pp.366–383. Available at: http://www.sciencedirect.com/science/article/pii/S0010218007000685 [Accessed November 22, 2014].

Diamantis, D.J., Kyritsis, D.C. & Goussis, D. a., 2015. The reactions supporting or opposing the development of explosive modes: Auto-ignition of a homogeneous methane/air mixture. *Proceedings of the Combustion Institute*, 35(1), pp.267–274. Available at: http://linkinghub.elsevier.com/retrieve/pii/S1540748914003733 [Accessed October 12, 2015].

Dixit, M. et al., 2015. Oxidative activation of methane on lanthanum oxide and nickel-lanthanum oxide catalysts. *Reaction Kinetics Mechanism and Catalysis*.

Fukada, S., 2004. Effects of temperature, oxygen-to-methane molar ratio and superficial gas velocity on partial oxidation of methane for hydrogen production. *International Journal of Hydrogen Energy*, 29(6), pp.619–625. Available at: http://www.sciencedirect.com/science/article/pii/S0360319903002258 [Accessed May 7, 2014].

Gallagher, S.M. et al., 2008. A rapid compression machine study of the oxidation of propane in the negative temperature coefficient regime. *Combustion and Flame*, 153(1-2), pp.316–333. Available at: http://linkinghub.elsevier.com/retrieve/pii/S0010218007002532 [Accessed October 11, 2015].

Goodwin, D.G., Moffat, H.K. & Speth, R.L., 2015. Cantera: An object- oriented software toolkit for chemical kinetics, thermodynamics, and transport processes. Available at: http://www.cantera.org.

Hartmann, L., Lucka, K. & Köhne, H., 2003. Mixture preparation by cool flames for diesel-reforming technologies. *Journal of Power Sources*, 118(1-2), pp.286–297. Available at: http://linkinghub.elsevier.com/retrieve/pii/S0378775303001009 [Accessed October 12, 2015].

Healy, D. et al., 2010. Methane/n-Butane Ignition Delay Measurements at High Pressure and Detailed Chemical Kinetic Simulations. *Energy & Fuels*, 24(3), pp.1617–1627. Available at: http://dx.doi.org/10.1021/ef901292j.

Healy, D. et al., 2008. Methane/propane mixture oxidation at high pressures and at high, intermediate and low temperatures. *Combustion and Flame*, 155(3), pp.451–461. Available at: http://linkinghub.elsevier.com/retrieve/pii/S001021800800196X [Accessed October 12, 2015].

Hoang, D.L. & Chan, S.H., 2004. Modeling of a catalytic autothermal methane reformer for fuel cell applications. *Applied Catalysis A: General*, 268(1-2), pp.207–216. Available at: http://www.sciencedirect.com/science/article/pii/S0926860X04002376 [Accessed September 2, 2014].

Huang, J. et al., 2004. Shock-tube study of methane ignition under engine-relevant conditions:

experiments and modeling. *Combustion and Flame*, 136(1-2), pp.25–42. Available at: http://linkinghub.elsevier.com/retrieve/pii/S0010218003002190 [Accessed October 12, 2015].

Joannon, M.D.E. et al., 2002. Dependence of Autoignition Delay on Oxygen Concentration in Mild Combustion of High Molecular Weight Paraffin. *Proceedings of the Combustion Institute*, 29, pp.1139–1146.

Mhadeshwar, A.B. & Vlachos, D.G., 2005. Is the water–gas shift reaction on Pt simple? *Catalysis Today*, 105(1), pp.162–172. Available at: http://www.sciencedirect.com/science/article/pii/S0920586105002075 [Accessed November 22, 2014].

Naidja, a, 2003. Cool flame partial oxidation and its role in combustion and reforming of fuels for fuel cell systems. *Progress in Energy and Combustion Science*, 29(2), pp.155–191. Available at: http://linkinghub.elsevier.com/retrieve/pii/S0360128503000182 [Accessed October 12, 2015].

Pekalski, A.A. et al., 2002. The relation of cool flames and auto-ignition phenomena to process safety at elevated pressure and temperature. , 93, pp.93–105.

Picarelli, A. et al., 2010. Auto-ignition delay times of methane / air diluted mixtures. Numerical and experimental approaches. In *Processes and Technologies for a Sustainable Energy*. pp. 1–8.

Prince, J.C., Williams, F. a. & Ovando, G.E., 2015. A short mechanism for the low-temperature ignition of n-heptane at high pressures. *Fuel*, 149(X), pp.138–142. Available at: http://linkinghub.elsevier.com/retrieve/pii/S0016236114008412 [Accessed October 12, 2015].

Ranzi, E. et al., 2007. A Wide Range Modeling Study of Methane Oxidation. *Combustion Science and Technology*, 96(April 2013), pp.37–41.

Ranzi, E. et al., 2010. CRECK modeling. Available at: http://creckmodeling.chem.polimi.it/C1C31006.CKI.

Sabia, P. et al., 2015. Dynamic Behaviors in Methane MILD and Oxy-Fuel Combustion. Chemical Effect of CO 2. *Energy & Fuels*, 29(3), pp.1978–1986. Available at: http://pubs.acs.org/doi/abs/10.1021/ef501434y [Accessed October 12, 2015].

Sabia, P. et al., 2013. Methane auto-ignition delay times and oxidation regimes in MILD combustion at atmospheric pressure. *Combustion and Flame*, 160(1), pp.47–55. Available at: http://linkinghub.elsevier.com/retrieve/pii/S0010218012002763 [Accessed October 12, 2015].

Sabia, P. et al., 2012. Modeling Negative Temperature Coefficient region in methane oxidation. *Fuel*, 91(1), pp.238–245. Available at: http://linkinghub.elsevier.com/retrieve/pii/S0016236111004236 [Accessed October 12, 2015].

Seidel, L. et al., 2015. Comprehensive kinetic modeling and experimental study of a fuel-rich, premixed n-heptane flame. *Combustion and Flame*, 162(5), pp.2045–2058. Available at: http://linkinghub.elsevier.com/retrieve/pii/S001021801500005X [Accessed September 9, 2015].

Smith, G.P. et al., GRI 30 Mechanism. Available at: http://www.me.berkeley.edu/gri_mech/.

U.S. Energy Information Administration, 2014. Annual Energy Outlook 2014. *Doe/Eia*, 0383, pp.1–269.

Vourliotakis, G., Skevis, G. & Founti, M. a., 2012. Combustion chemistry aspects of alternative fuels reforming for high-temperature fuel cell applications. *International Journal of Hydrogen Energy*, 37(21), pp.16649–16662. Available at: http://linkinghub.elsevier.com/retrieve/pii/S0360319912004788 [Accessed September 9, 2015].

Vourliotakis, G., Skevis, G. & Founti, M.A., 2009. Detailed kinetic modelling of non-catalytic ethanol reforming for SOFC applications. *International Journal of Hydrogen Energy*, 34(18), pp.7626–7637. Available at: http://www.sciencedirect.com/science/article/pii/S0360319909010301 [Accessed January 1, 2015].

Wehinger, G.D., Eppinger, T. & Kraume, M., 2015. Detailed numerical simulations of catalytic fixed-bed reactors: Heterogeneous dry reforming of methane. *Chemical Engineering Science*, 122, pp.197–209. Available at: http://www.sciencedirect.com/science/article/pii/S0009250914005016 [Accessed November 21, 2014].

Wei, J., 2004. Structural requirements and reaction pathways in methane activation and chemical conversion catalyzed by rhodium. *Journal of Catalysis*, 225(1), pp.116–127. Available at: http://linkinghub.elsevier.com/retrieve/pii/S0021951704001836 [Accessed June 10, 2015].

Xu, Y. et al., 2014. Numerical simulation of natural gas non-catalytic partial oxidation reformer. *International Journal of Hydrogen Energy*, 39(17), pp.9149–9157. Available at: http://www.sciencedirect.com/science/article/pii/S036031991400932X [Accessed January 1, 2015].

Zádor, J., Taatjes, C. a. & Fernandes, R.X., 2011. Kinetics of elementary reactions in low-temperature autoignition chemistry. *Progress in Energy and Combustion Science*, 37(4), pp.371–421. Available at: http://linkinghub.elsevier.com/retrieve/pii/S0360128510000559 [Accessed October 12, 2015].

Zhou, X., Chen, C. & Wang, F., 2010. Modeling of non-catalytic partial oxidation of natural gas under conditions found in industrial reformers. *Chemical Engineering and Processing: Process Intensification*, 49(1), pp.59–64. Available at: http://www.sciencedirect.com/science/article/pii/S0255270109002402 [Accessed January 1, 2015].